\pdfoutput=1
\RequirePackage{ifpdf}
\ifpdf 
\documentclass[pdftex]{sigma}
\else
\documentclass{sigma}
\fi

\numberwithin{equation}{section}

{\theoremstyle{definition} \newtheorem{notation}{Notation}

}

\numberwithin{proposition}{section}
\numberwithin{remark}{section}
\numberwithin{notation}{section}

\begin{document}

\allowdisplaybreaks

\renewcommand{\thefootnote}{$\star$}

\renewcommand{\PaperNumber}{046}

\FirstPageHeading

\ShortArticleName{Another New Solvable Many-Body Model of Goldf\/ish Type}

\ArticleName{Another New Solvable Many-Body Model\\ of Goldf\/ish Type\footnote{This
paper is a contribution to the Special Issue ``Geometrical Methods in Mathematical Physics''. The full collection is available at \href{http://www.emis.de/journals/SIGMA/GMMP2012.html}{http://www.emis.de/journals/SIGMA/GMMP2012.html}}}

\Author{Francesco CALOGERO}

\AuthorNameForHeading{F.~Calogero}

\Address{Physics Department, University of Rome ``La Sapienza'',\\ Istituto Nazionale di Fisica Nucleare, Sezione di Roma, Italy}
\Email{\href{francesco.calogero@roma1.infn.it}{francesco.calogero@roma1.infn.it}, \href{francesco.calogero@uniroma1.it}{francesco.calogero@uniroma1.it}}

\ArticleDates{Received May 03, 2012, in f\/inal form July 17, 2012; Published online July 20, 2012}

\Abstract{A new \textit{solvable} many-body problem is identif\/ied.
It is characterized by nonlinear \textit{Newtonian} equations of motion
(``acceleration equal force'') featuring one-body and two-body
velocity-dependent forces ``of goldf\/ish type'' which determine the motion of
an arbitrary number $N$ of unit-mass point-particles in a plane. The $N$
(generally \textit{complex}) values $z_{n}( t) $ at time $t$ of
the $N$ coordinates of these moving particles are given by the $N$
eigenvalues of a~time-dependent $N\times N$ matrix $U( t) $
explicitly known in terms of the $2N$ initial data $z_{n}( 0) $
and $\dot{z}_{n}(0) $. This model comes in two dif\/ferent
variants, one featuring 3 arbitrary coupling constants, the other only~2;
for special values of these parameters \textit{all} solutions are completely
periodic with the same period independent of the initial data (``\textit{isochrony}''); for other special values of these parameters this property
holds up to corrections vanishing exponentially as $t\rightarrow \infty $ (``\textit{asymptotic isochrony}''). Other \textit{isochronous} variants of
these models are also reported. Alternative formulations, obtained by
changing the dependent variables from the $N$ zeros of a monic polynomial of
degree~$N$ to its~$N$ coef\/f\/icients, are also exhibited. Some mathematical
f\/indings implied by some of these results~-- such as \textit{Diophantine}
properties of the zeros of certain polynomials~-- are outlined, but their
analysis is postponed to a separate paper.}


\Classification{37J35; 37C27; 70F10; 70H08}

\renewcommand{\thefootnote}{\arabic{footnote}}
\setcounter{footnote}{0}

\section{Introduction}\label{section1}

The technique used in this paper to identify a new \textit{solvable}
many-body problem has become by now standard. Its more convenient version
starts from the identif\/ication of a \textit{solvable} matrix problem
characterized by two \textit{first-order}, generally \textit{autonomous},
matrix ODEs def\/ining the time evolution of two $N\times N$ matrices $U\equiv
U ( t ) $ and $V\equiv V ( t ) $:
\begin{subequations}
\begin{gather}
\dot{U}=F ( U,V ),\qquad \dot{V}=G ( U,V) ,  \label{UVdot}
\end{gather}
where the two functions $F( U,V)$, $G( U,V) $ may
depend on several scalar parameters but on no other matrix besides~$U$ and~$V$. Here and hereafter superimposed dots denote of course dif\/ferentiations
with respect to the independent variable $t$ (``time''). The \textit{solvable}
character of this system amounts to the possibility to obtain the solution
of its initial-value problem,
\begin{gather}
U ( t ) =U ( t;U_{0},V_{0} )  ,\qquad V ( t )
=V ( t;U_{0},V_{0} ) ,\qquad U_{0}\equiv U ( 0 )
 ,\qquad U_{0}\equiv V ( 0 )  ,
\end{gather}
\end{subequations}
with the two matrix functions $U ( t;U_{0},V_{0} )$, $V (t;U_{0},V_{0} ) $ \textit{explicitly} known. For instance, in this
paper we shall focus (see~(\ref{EqW}) below) on a system~(\ref{UVdot}) which
can be related to a \textit{linear} (third-order, matrix) evolution equation
(see~(\ref{EqW}) below), so that its initial-value problem can be \textit{explicitly solved}.

One then introduces the eigenvalues $z_{n} ( t ) $ of one of these
two matrices, by setting, say
\begin{gather}
U ( t ) =R ( t )  Z ( t )   [ R (
t )  ] ^{-1} ,\qquad Z ( t ) =\text{diag} [ z_{n} (t)] .  \label{UZ}
\end{gather}

\begin{remark}\label{remark1.1}
The diagonalizing matrix $R(t) $ is
identif\/ied by this formula up to right-mul\-ti\-pli\-ca\-tion by an \textit{%
arbitrary diagonal} $N\times N$ matrix $D ( t )$, $R ( t )
\Rightarrow \tilde{R} ( t ) \equiv R ( t )  D (t) $.
\end{remark}

Then one introduces a new $N\times N$ matrix $Y( t) $ by setting
\begin{subequations}\label{V}
\begin{gather}
V(t) =R(t)  Y(t)   [ R ( t )  ] ^{-1} ,\qquad Y(t) = [ R(t)  ]
^{-1} V(t)  R(t)  ,  \label{VYR}
\end{gather}
where of course the $N\times N$ matrix $R(t)$ is that def\/ined
above, see (\ref{UZ}). This matrix $Y(t)$ is of course
generally \textit{nondiagonal}:
\begin{gather}
Y_{nm}(t) =\delta _{nm} y_{n}(t) + ( 1-\delta_{nm}) Y_{nm}(t) .  \label{Y}
\end{gather}
\end{subequations}

\begin{notation}\label{notation1.1}
Indices such as $n$, $m$, $\ell $ run from $1$ to $N$
(unless otherwise explicitly indicated), and $N$ is an \textit{arbitrary
positive integer} (indeed generally $N\geq 2$). $\delta _{nm}$ is the
Kronecker symbol, $\delta _{nm}=1$ if $n=m$, $\delta _{nm}=0$ if $n\neq m$.
Note that hereafter we use the notation $Y_{nm}$ to denote the $N\left(
N-1\right) $ \textit{off-diagonal} elements of the $N\times N$ matrix~$Y$.
\end{notation}

It often turns out that the time evolution of the eigenvalues $z_{n}\left(
t\right) $ is then characterized by a~system of $N$ second-order ODEs which
read as follows:
\begin{gather}
\ddot{z}_{n}=f ( z_{n},\dot{z}_{n} ) +\sum_{\ell =1,\, \ell =n}^{N}
\left[ Y_{n\ell } Y_{\ell n} \frac{g^{ ( 1 ) } ( z_{n},\dot{z}
_{n} )  g^{ ( 2 ) } ( z_{\ell },\dot{z}_{\ell } ) }{z_{n}-z_{\ell }}\right]  ,  \label{zndotdot}
\end{gather}
where the three functions $f ( z,\dot{z} )$, $g^{ ( 1 )
} ( z,\dot{z} ) $ and $g^{ ( 2 ) } ( z,\dot{z} ) $
can be computed from the two matrix functions $F ( U,V )$, $G (
U,V ) $ (see below). It is then natural to try and interpret this
system of ODEs as the \textit{Newtonian} equations of motion (``acceleration
equal force'') of a many-body problem characterizing the motion of $N$
particles whose coordinates coincide with the $N$ eigenva\-lues~$z_{n} (
t ) $; an $N$-body problem generally featuring one-body and two-body
velocity-dependent forces, with the $N ( N-1 ) $ quantities $Y_{n\ell } Y_{\ell n}$ playing the role of ``coupling constants''. But these
quantities are not time-independent, nor can they be arbitrarily assigned:
they are the \textit{off-diagonal} elements of the $N\times N$ matrix~$Y$,
hence they should be themselves considered as \textit{dependent variables},
the time evolution of which is characterized by the system of $N(
N-1) $ ODEs implied for them by (\ref{UVdot}) via (\ref{UZ}) and (\ref{V}).

Two options are then open to provide nonetheless a ``physical'' interpretation
for the equations of motion (\ref{zndotdot}).

One option that we do not pursue here is to provide some kind of ``physical''
interpretation for these $N ( N-1 ) $ quantities $Y_{n\ell
} Y_{\ell n}$ as additional (internal) degrees of freedom of the moving
particles.

A second option~-- the one which we pursue below~-- is to f\/ind a
(time-independent) \textit{ansatz} expressing the $N ( N-1 ) $
quantities $Y_{n\ell }$ in terms of the $N$ coordinates $z_{m},$ or possibly
of the $2N$ quantities $z_{m}$, $\dot{z}_{m}$; an \textit{ansatz} consistent
with the $N ( N-1 ) $ equations of motion satisf\/ied by the $N (
N-1 ) $ quantities $Y_{n\ell }$, which satisf\/ies these equations either
\textit{identically} (i.e., independently from the time evolution of the $N$
coordinates $z_{m}(t) $) or, as it were, \textit{self-consistently} (i.e., thanks to the time evolution~(\ref{zndotdot}) of
the $N$ coordinates $z_{m}(t) $ with the $N ( N-1 ) $
quantities $Y_{n\ell }$ assigned according to the \textit{ansatz}). Given a
matrix system of type~(\ref{UVdot}) no technique is known to assess \textit{%
a priori} whether or not such an \textit{ansatz} exist. However the
experience accumulated over time suggests that, if such an \textit{ansatz}
does exist, it has one of the following two forms:
\begin{subequations}\label{Ans}
\begin{gather}
\text{ansatz 1}:\quad Y_{n\ell }=\frac{g^{ ( 1 )
} ( z_{n} )  g^{ ( 2 ) } ( z_{\ell } ) }{z_{n}-z_{\ell }} ;  \label{Ans1}
\\
\text{ansatz 2}:\quad Y_{n\ell }=\left\{ g^{ ( 1 )
} ( z_{n} )  g^{ ( 2 ) } ( z_{\ell } )  \big[
\dot{z}_{n}+f^{ ( 1 ) } ( z_{n} ) \big] \big[ \dot{z}
_{\ell }+f^{ ( 2 ) } ( z_{\ell } ) \big] \right\}
^{1/2};  \label{Ans2}
\end{gather}
\end{subequations}
of course with the functions appearing in the right-hand side of these
formulas chosen appropriately. And as a rule the \textit{ansatz~1} should
work \textit{identically}, i.e.\ independently of the time evolution of the
coordinates $z_{m}(t) $; while to ascertain the validity of
\textit{ansatz~2} the equations of motion~(\ref{zndotdot}) should be used
\textit{self-consistently}.

For instance in the very simple case of the equations of motion (\ref{UVdot}) with $F( U,V) =V$ and $G( U,V) =0$ implying $\ddot{U}
=0$ and $U(t) =U_{0}+V_{0} t$, $V(t) =V_{0}$, both
\textit{ans\"{a}tze} exist: the \textit{ansatz~1} reads in this case $%
Y_{n\ell }=i g/ ( z_{n}-z_{\ell } ) $ with $i$ the imaginary unit
(introduced for convenience) and $g$ an \textit{arbitrary constant}, and it
yields the prototypical ``CM'' model characterized by the equations of motion
\begin{gather}
\ddot{z}_{n}=2 g^{2}\sum_{\ell =1,\, \ell \neq n}^{N} ( z_{n}-z_{\ell
} ) ^{-3} ;  \label{CalMos}
\end{gather}
while the \textit{ansatz~2} in this case reads $Y_{n\ell }= ( \dot{z}%
_{n} \dot{z}_{\ell } ) ^{1/2},$ and it yields the prototypical
``goldf\/ish'' model characterized by the equations of motion
\begin{subequations}
\begin{gather}
\ddot{z}_{n}=\sum_{\ell =1,\, \ell \neq n}^{N}\left( \frac{2 \dot{z}_{n} \dot{z}
_{\ell }}{z_{n}-z_{\ell }}\right) .  \label{Gold}
\end{gather}

{\bf Nomenclature and historical remarks}. The model characterized by the
Newtonian equations of motion~(\ref{CalMos})~-- which obtain from the
Hamiltonian
\[
H_{\rm CM} ( \underline{z}, \underline{p} ) =\frac 12 \sum_{n=1}^{N}p_{n}^{2}+\frac 12  g^{2}\sum_{n,m=1,\, n\neq
m}^{N} ( z_{n}-z_{m} ) ^{-2}
\]
--~is usually associated with the
names of those who f\/irst demonstrated the possibility to treat this
many-body problem exactly, respectively in the \textit{quantal} \cite{C1971}
and in the \textit{classical} \cite{M1975} contexts; accordingly, we usually
call ``many-body problems of CM-type'' those featuring in the right-hand
(``forces'') side of their Newtonian equations of motion a term such as that
appearing in the right-hand side of~(\ref{CalMos}) (in addition of course to
other terms). Such models are typically produced by the \textit{ansatz~1}.

The \textit{solvable} character of the many-body problem characterized by
the Newtonian equations of motion (\ref{Gold})~-- which is also Hamiltonian,
for instance with the Hamiltonian~\cite{C1996, CF1997,C2008}
\[
 H_{\rm gold} ( \underline{z}, \underline{p} ) =\sum_{n=1}^{N}\left[ \exp  ( p_{n} )
 \prod\limits_{m=1,\, m\neq n}^{N} ( z_{n}-z_{m} ) ^{-1}\right]
 \]
-- is demonstrated by the following
neat \textit{Prescription} \cite{C2008, C1978,C1999}: the $N$
values of the coordina\-tes~$z_{n}(t) $  providing the solution of
the initial-value problem of the equations of motion~(\ref{Gold}) are the~$N$
\ zeros of the following algebraic equation for the variable $z$:
\begin{gather}
\sum_{n=1}^{N}\left[ \frac{\dot{z}_{n} ( 0 ) }{z-z_{n} (
0 ) }\right] =\frac{1}{t} .  \label{SolGold}
\end{gather}
\end{subequations}
(Note that this formula amounts to a polynomial equation of degree $N$ in~$z $, as it is immediately seen by multiplying it by the polynomial $%
\prod\limits_{m=1}^{N} [ z-z_{m} ( 0 )  ] $). In \cite{C1999} it was suggested that this model, in view of its neat character, be
considered a ``goldf\/ish'' (meaning, in Russian traditional lore, a very
remarkable item, endowed with magical properties); accordingly, we usually
call ``many-body problems of goldf\/ish-type'' those featuring in the right-hand
(``forces'') side of their Newtonian equations of motion a term such as that
appearing in the right-hand side of~(\ref{Gold}) (in addition of course to
other terms). Such models are typically produced by the \textit{ansatz~2}.

The original idea of the approach described above is due to Olshanetsky and
Perelomov, who introduced it to solve, in the classical context, the
many-body model characterized by the Newtonian equations of motion (\ref{CalMos})~\cite{OP1976}. For a more detailed description of their work see
the review paper~\cite{OP1981}, the book~\cite{P1990}, Section~2.1.3.2
(entitled ``The technique of solution of Olshanetsky and Perelomov'') in \cite{C2001}, and other references cited in these books. In Section~4.2.2
(entitled ``Goldf\/ishing'') of~\cite{C2008} several many-body problems, mainly
``of goldf\/ish type'', are reviewed, the~\textit{solvable} character of which
has been ascertained by this approach; and several other such models are
discussed in more recent papers~\cite{C2010, C2011,C2012,C2012a}.

\looseness=-1
The present paper provides two further additions to the list of \textit{solvable} many-body problems ``of goldf\/ish type''; and we expect that other
items will be added to this list in the future, possibly by a continuation
of the case-by-case search of \textit{solvable} matrix evolution equations
allowing~-- via the route outlined above and described in more detail, in a
specif\/ic case, below (see Section~\ref{section3})~-- the identif\/ication of working \textit{ans\"{a}tze} leading to \textit{new} systems of Newtonian equations (thereby
shown to be themselves \textit{solvable}, inasmuch as their solution is
reduced to the \textit{algebraic} task of computing the $N$ eigenvalues of
an explicitly known $N\times N$ matrix). The identif\/ication of a \textit{new}
model of this kind constitutes~-- in our opinion~-- an interesting f\/inding
(even if several analogous models have been previously discovered): we view
these many-body problems as gems embedded in the magma of the generic
many-body problems which are not amenable to exact treatments (although the
latter include of course more examples of applicative interest and are also
mathe\-matically interesting to investigate the emergence and phenomenology of
chaotic behaviors).

In the following Section~\ref{section2} the main f\/indings of this paper are reported,
including in particular a description of two new \textit{solvable} many-body
problems (one featuring~3, the other only~2, \textit{a~priori arbitrary} parameters), of the algebraic solution of their initial-value
problems, and of the variety of behaviors (including \textit{isochrony} and
\textit{asymptotic isochrony}) featured by them for certain assignments of
their parameters; two additional \textit{isochronous} systems are moreover
exhibited in Section~\ref{section2.1}. In Section~\ref{section3} these results are proven. Section~\ref{section4}
provides the alternative formulations of these models, obtained by changing
the dependent variables from the $N$ zeros of a~monic polynomial of degree $N $ to its $N$ coef\/f\/icients. A f\/inal Section~\ref{section5} entitled ``Outlook'' outlines
further developments, including in particular the identif\/ication of \textit{Diophantine} properties of the zeros of certain polynomials; but their
detailed discussion is postponed to a separate paper.

\section{Main f\/indings}\label{section2}

The two models treated in this paper are characterized by the following two
sets of \textit{Newtonian} equations of motion ``of goldf\/ish type''.

\textit{Model $(i)$}:
\begin{subequations}\label{NewGold}
\begin{gather}
\ddot{z}_{n}  = -3 \dot{z}_{n} z_{n}+\gamma  \dot{z}_{n}-z_{n}^{3}+\gamma
 z_{n}^{2}+ [ -a+b  ( \gamma +b )  ]  z_{n}+a (
\gamma +b )  \notag \\
\hphantom{\ddot{z}_{n}  =}{} +\sum_{\ell =1,\, \ell \neq n}^{N}\left[ \frac{2 \big( \dot{z}
_{n}+a+b z_{n}+z_{n}^{2} \big)  \big( \dot{z}_{\ell }+a+b z_{\ell
}+z_{\ell }^{2} \big) }{z_{n}-z_{\ell }}\right]  ,  \label{NewGold1}
\end{gather}
where $a$, $b$ and $\gamma $ are 3 \textit{a priori arbitrary} parameters.

\textit{Model $(ii)$}:
\begin{gather}
\ddot{z}_{n}=-3 \dot{z}_{n} z_{n}-3 b \dot{z}_{n}-z_{n}^{3}-3 b z_{n}^{2}-2
\big( a+b^{2}\big)  z_{n}-2 a b    \notag \\
\hphantom{\ddot{z}_{n}=}{}
+\sum_{\ell =1,\, \ell \neq n}^{N}\left[ \frac{2 \big( \dot{z}
_{n}+a+b z_{n}+z_{n}^{2}/2 \big)  \big( \dot{z}_{\ell }+a+b z_{\ell
}+z_{\ell }^{2}/2 \big) }{z_{n}-z_{\ell }}\right]  ,    \label{NewGold2}
\end{gather}
\end{subequations}
where $a$ and $b$ are 2 \textit{a priori arbitrary} parameters.

\begin{notation}\label{notation2.1}
 These models describe the motion of an \textit{arbitrary} number $N$ (generally $N\geq 2$) of points moving in the \textit{complex} $z$-plane (but see below Remark~\ref{remark2.1}). Their positions are
identif\/ied by the \textit{complex} dependent variables $z_{n}\equiv
z_{n}(t)$. The independent variable $t$ (``time'') is instead
\textit{real}. Superimposed dots denote of course time-dif\/ferentiations. The
parameters featured by these models are generally \textit{arbitrary complex}
numbers; unless otherwise specif\/ied when discussing special cases.
\end{notation}

\begin{remark}\label{remark2.1}
 It is possible to reformulate these models so that they
describe the motion of pointlike ``physical particles'' moving in a \textit{real}~-- say, horizontal~-- plane: see for instance, in~\cite{C2001}, Section~4.1 entitled ``How to obtain by complexif\/ication rotation-invariant many-body
models in the plane from certain many-body models on the line''. This task is
left to the interested reader. But hereafter we feel free to refer to the
models identif\/ied by the \textit{Newtonian} equations of motion (\ref%
{NewGold}) as many-body problems characterizing the motion of $N$ particles
in a plane.
\end{remark}

\begin{remark}\label{remark2.2}
  Additional parameters could be inserted in these models
by shifting or rescaling the dependent variables $z_{n}$ or the independent
variable $t$. We will not indulge in such trivial exercises (see also Remark~\ref{remark3.1} below).
\end{remark}

The \textit{solvable} character of these two models is demonstrated by the
following

\begin{proposition}\label{proposition2.1}
The solution of the initial-value problems of the
two many-body models cha\-rac\-terized by the Newtonian equations of motion ``of
goldfish type'' \eqref{NewGold} are given by the eigenvalues of the $N\times
N $ matrix $U(t)$, the explicit expression of which in terms of
the $2N$ initial data~$z_{n} ( 0 )$,~$\dot{z}_{n} ( 0 ) $
and of the time $t$ is given by the following formulas $($see \eqref{GenSolU}
with \eqref{SolAB}$)$:
\begin{subequations}\label{SolU}
\begin{gather}
U(t) =i \big\{ I+A \exp  [ i  ( \omega _{2}-\omega
_{1} )  t ] +B \exp  [ i  ( \omega _{3}-\omega _{1} )
 t ] \big\} ^{-1}   \notag \\
\hphantom{U(t) =}{}
\times \big\{ \omega _{1} I+\omega _{2} A \exp  [ i  ( \omega
_{2}-\omega _{1} )  t ] +\omega _{3} B \exp  [ i  ( \omega
_{3}-\omega _{1} )  t ] \big\}  ,
\end{gather}
with the two \textit{constant} $N\times N$ matrices $A$ and $B$ defined as
follows:
\begin{gather}
A  = - ( \omega _{1}-\omega _{3} )   ( \omega _{2}-\omega
_{3} ) ^{-1} \big[ I-i  ( \omega _{1}+\omega _{3} )  \big(
V_{0}+\omega _{3}^{2}\big) ^{-1}  ( U_{0}-i \omega _{3} ) ^{-1}
\big]   \notag \\
\hphantom{A=}{}\times
 \big[ I-i  ( \omega _{2}+\omega _{3} )  \big(
V_{0}+\omega _{3}^{2}\big) ^{-1}  ( U_{0}-i \omega _{3} ) ^{-1}
\big] ^{-1},
\\
B  =  ( \omega _{1}-\omega _{2} )   ( \omega _{2}-\omega
_{3} ) ^{-1} \big[ I-i  ( \omega _{1}+\omega _{2} )  \big(
V_{0}+\omega _{2}^{2}\big) ^{-1}  ( U_{0}-i \omega _{2} ) ^{-1}
\big]    \notag \\
\hphantom{B=}{}\times
 \big[ I-i  ( \omega _{2}+\omega _{3} )  \big(
V_{0}+\omega _{2}^{2}\big) ^{-1}  ( U_{0}-i \omega _{2} ) ^{-1}
\big] ^{-1} .
\end{gather}
\end{subequations}
Here and hereafter $I$ is the $N\times N$ \textit{unit} matrix, the $N\times
N$ matrix $U_{0}\equiv U ( 0 ) $ is \textit{diagonal} and is given
in terms of the \textit{initial} particle positions $z_{n}(0)$
as follows,
\begin{gather}
U_{0}=\text{\rm diag} [ z_{n}(0) ] ,  \label{Uzero}
\end{gather}
while the $N\times N$ matrix $V_{0}\equiv V(0) $ is the sum of a
\textit{diagonal} and a \textit{dyadic} matrix, being given componentwise by
the following formulas in terms of the \textit{initial} particle positions $%
z_{n}(0) $ and velocities $\dot{z}_{n}(0) $:
\begin{subequations}\label{Vzero}
\begin{gather}
( V_{0}) _{nm}=-\delta _{nm} \big[ a+bz_{n}(0)
+( c-1) z_{n}^{2}(0) \big] +V_{n}V_{m},
\\
V_{n}=\big[ \dot{z}_{n}(0) +a+b z_{n}(0)
+c z_{n}^{2}(0) \big] ^{1/2},
\end{gather}
\end{subequations}
with $c=1$ for \textit{model $(i)$} and $c=1/2$ for \textit{model $(ii)$} $($see \eqref{cc}$)$. As for the $3$ constants $\omega _{j}$ appearing in \eqref{SolU},
they are defined by the following formula $($see \eqref{omega123}$)$:
\begin{subequations}\label{omegaj}
\begin{gather}
\omega ^{3}+i \gamma  \omega ^{2}+\beta   \omega -i \alpha = ( \omega
-\omega _{1} )  ( \omega -\omega _{2} )  ( \omega -\omega
_{3} )
\end{gather}
implying
\begin{gather}
\alpha =-i\omega _{1}\omega _{2}\omega _{3}, \qquad \beta =\omega _{1}\omega
_{2}+\omega _{2}\omega _{3}+\omega _{3}\omega _{1}, \qquad \gamma =i(
\omega _{1}+\omega _{2}+\omega _{3}),  \label{alphabetagammaomega}
\end{gather}
\end{subequations}
with $\alpha $ and $\beta $ respectively $\alpha$, $\beta $ and $\gamma $
given by the following expressions for \textit{model $(i)$} respectively for
\textit{model $(ii)$} $($see \eqref{alphabeta1} respectively \eqref{alphabetagamma2}$)$:

\textit{model $(i)$}:
\begin{subequations}
\label{cc}
\begin{gather}
c=1,\qquad \alpha =a ( \gamma +b ) ,\qquad \beta =-a+b ( \gamma
+b )  ;
\end{gather}

\textit{model $(ii)$}:
\begin{gather}
c=\frac{1}{2} ,\qquad \alpha =-2 a b, \qquad \beta =-2 \big( a+b^{2}\big),\qquad \gamma =-3b.
\end{gather}
\end{subequations}
\end{proposition}

Note that the 3 \textit{a priori arbitrary} parameters $\omega _{j}$ have
the dimension of an \textit{inverse time}; above and hereafter we assume for
simplicity that they are \textit{different} among each other (except in the
following Subsection~\ref{section2.1}, where they are all assumed to vanish).

\begin{subequations}\label{periodic}
It is plain (see (\ref{SolU})) that, if the 3 constants $\omega _{j}$ are
\textit{integer} multiples of a single \textit{real} cons\-tant~$\omega $,
\begin{gather}
\omega _{j}=k_{j} \omega ,\qquad j=1,2,3,  \label{omegaintegers}
\end{gather}%
then the matrix $U(t) $ is periodic,
\begin{gather}
U(t) =U ( t\pm T ) ,  \label{Uperiodic}
\end{gather}
\end{subequations}
with period
\begin{gather}
T=\frac{2\pi }{ \vert \omega  \vert }.  \label{T}
\end{gather}
Here and throughout the 3 parameters $k_{j}$ are \textit{integer numbers}
(positive, negative or vanishing, but dif\/ferent among themselves); their
def\/inition, as well as that of the \textit{positive} parameter $\omega $, is
made unequivocal (up to permutations; once the 3 parameters $\alpha$, $\beta$, $\gamma $ are assigned, compatibly via~(\ref{alphabetagammaomega}) with~(\ref{omegaintegers})) by the requirement that this \textit{positive}
parameter~$\omega $ be assigned the \textit{largest} value for which (\ref{omegaintegers}) holds.

More generally, if the \textit{real parts} of the 3 constants $\omega _{j}$
are \textit{integer} multiples of a single \textit{real} constant $\omega $
and the \textit{imaginary parts} of 2 of them coincide while the \textit{imaginary part} of the third is larger, say,
\begin{subequations}
\begin{gather}
{\rm Re} ( \omega _{j} ) =k_{j}\omega ,\quad j=1,2,3;\qquad {\rm Im}( \omega _{1}) ={\rm Im}( \omega _{2}) <{\rm Im}( \omega _{3}),  \label{Reomegaintegers}
\end{gather}%
then the matrix $U(t) $ is \textit{asymptotically periodic} with
period $T$, namely it becomes periodic with period $T$ in the remote future
\textit{up to exponentially vanishing corrections}, so that
\begin{gather}
\lim_{t\rightarrow \infty } \vert U(t) -U (
t\pm T )  \vert =0 .  \label{Uasyperiodic}
\end{gather}
\end{subequations}

While for \textit{generic} values of the parameters, implying via (\ref{omegaj}) that the \textit{imaginary parts} of the $3$ quantities $\omega
_{j}$ are \textit{different} among themselves,
\begin{subequations}
\label{Imomegajdiff}
\begin{gather}
{\rm Im} ( \omega _{1} ) \neq {\rm Im} ( \omega _{2} )
 ,\qquad {\rm Im} ( \omega _{2} ) \neq {\rm Im} ( \omega
_{3} )  ,\qquad {\rm Im} ( \omega _{3} ) \neq {\rm Im} (\omega _{1} ) ,  \label{Imomegadiff}
\end{gather}
then clearly $U(t) $ tends to a time-independent matrix as $t\rightarrow \pm \infty $,
\begin{gather}
U(t) \underset{t\rightarrow \pm \infty }{\rightarrow }U(\pm \infty ).  \label{Uasy}
\end{gather}
\end{subequations}

Let us emphasize that these outcomes, (\ref{Uperiodic}) or (\ref{Uasyperiodic}) or (\ref{Uasy}), obtain~-- provided the 3~parameters $\alpha$,
$\beta$, $\gamma $ satisfy~(\ref{alphabetagammaomega}) with~(\ref{omegaintegers}) or~(\ref{Reomegaintegers}) or~(\ref{Imomegadiff})~-- for
\textit{arbitrary initial data}, and that the period $T$ is as well \textit{%
independent of the initial data}. Let us also note that these properties
hold unless $U(t) $ is \textit{singular}; clearly a \textit{nongeneric} circumstance, in which case~$U(t) $ would in fact
still feature the properties indicated above, but only in the sense in
which, for instance, the function $\tan  [ \omega  ( t-t_{0} )
/2 ] $ (with~$\omega $ and~$t_{0}$ two \textit{real} numbers) is
periodic with period~$T$.

These properties of the $N\times N$ matrix $U(t) $ carry of
course over to its eigenvalues $z_{n}(t) $, hence to the generic
solutions of the many-body problems ``of goldf\/ish type''~(\ref{NewGold});
these models are therefore \textit{isochronous} respectively \textit{asymptotically isochronous} if their parameters satisfy the relevant
conditions, see~(\ref{alphabetagammaomega}) with~(\ref{omegaintegers})
respectively~(\ref{Reomegaintegers}).

\begin{remark}\label{remark2.3}
Let us however recall that the periods of the time
evolution of individual eigenvalues of a periodic matrix may be a (generally
small) \textit{positive integer multiple} of the period of the matrix, due
to the possibility that dif\/ferent eigenvalues exchange their roles over the
time evolution (for a discussion of this possibility see~\cite{GS2005},
where a justif\/ication is also provided of the statement made above that the
relevant \textit{positive integer multiple} is ``generally small'').
\end{remark}

\subsection{Two additional \textit{isochronous} many-body models}\label{section2.1}

Additional \textit{isochronous} models obtain by applying, to the special
cases of the two many-body models (\ref{NewGold}) with all parameters
vanishing,
\begin{gather}
\ddot{z}_{n}=-3 \dot{z}_{n} z_{n}-z_{n}^{3}++\sum_{\ell =1,\, \ell \neq n}^{N}
\left[ \frac{2  ( \dot{z}_{n}+c z_{n}^{2} )  ( \dot{z}_{\ell
}+c z_{\ell }^{2} ) }{z_{n}-z_{\ell }}\right] ,
\end{gather}
with $c=1$ respectively $c=1/2$, the standard ``isochronizing'' trick, see for
instance Section 2.1 (entitled ``The trick'') of~\cite{C2008}. It amounts in
these cases to the following change of dependent and independent variables,
\begin{subequations}
\begin{gather}
\tilde{z}_{n}(t) =\exp ( i\omega t) z_{n}(
\tau ) ,\qquad \tau =\frac{\exp  ( i\omega t) -1}{i\omega },  \label{ztildetau}
\end{gather}
implying
\begin{gather}
\tilde{z}_{n}(0) =z_{n}(0) ,\qquad \overset{\cdot }{\tilde{z}}_{n}(0) =z_{n}^{\prime }(0) +i\omega
z_{n}(0) .
\end{gather}
\end{subequations}
Here and below $\omega $ is again an arbitrary \textit{real} constant to
which we associate the period $T$, see~(\ref{T}), and of course appended
primes denote dif\/ferentiations with respect to the argument of the function
they are appended to (hence $z_{n}^{\prime }(\tau)
=dz_{n}(\tau) /d\tau $). Hence clearly the Newtonian equations
of motion of these two models read as follows:
\begin{gather}
\overset{\cdot \cdot }{\tilde{z}}_{n}  = 3 i \omega  \overset{\cdot }{\tilde{z}}_{n}
+2\omega ^{2}\tilde{z}_{n}-3\overset{\cdot }{\tilde{z}}_{n}\tilde{z}_{n}+3i\omega \tilde{z}_{n}^{2}-\tilde{z}_{n}^{3}  \notag \\
\hphantom{\overset{\cdot \cdot }{\tilde{z}}_{n}  =}{}
+\sum_{\ell =1,\, \ell \neq n}^{N}\left[ \frac{2  ( \overset{\cdot }{\tilde{z}}_{n}-i\omega \tilde{z}_{n}+c\tilde{z}_{n}^{2}) (
\overset{\cdot }{\tilde{z}}_{\ell }-i\omega \tilde{z}_{\ell }+c\tilde{z}_{\ell }^{2}) }{\tilde{z}_{n}-\tilde{z}_{\ell }}\right],
\label{NewIsoGold}
\end{gather}
again with $c=1$ respectively $c=1/2$. The solutions of the corresponding
initial-value problems are clearly given by the following variant (of the
case with $\omega _{j}=0$) of Proposition~\ref{proposition2.1}:

\begin{proposition}\label{proposition2.2}
The solution of the initial-value problems of the
two many-body models cha\-racterized by the Newtonian equations of motion ``of
goldfish type'' \eqref{NewIsoGold} are given by the eigenvalues of the $N\times N$ matrix $\tilde{U}(t) $ given by the following
formulas in terms of the $2N$ initial data~$\tilde{z}_{n}(0)$,~$\overset{\cdot }{\tilde{z}}_{n}(0) $ and of the time~$t$:
\begin{subequations}
\label{SolUtilde}
\begin{gather}
\tilde{U}(t) =\exp  ( i\omega  t ) \big( I+\tilde{A}\tau +\tilde{B}\tau ^{2}\big) ^{-1} ( \tilde{A}+2\tilde{B}\tau
) ,  \label{Utilde}
\end{gather}
with the two \textit{constant} $N\times N$ matrices $\tilde{A}$ and $\tilde{B}$ defined as follows:
\begin{gather}
\tilde{A}=\text{\rm diag} [ \tilde{z}_{n}(0)  ]  ,\qquad
\tilde{B}_{nm} =\frac{1}{2} v_{n} v_{m},\qquad
v_{n}=\overset{\cdot }{\tilde{z}}_{n} (0) -i\omega \tilde{z}_{n}(0) +c\tilde{z}_{n}^{2}(0),  \label{ABtilde}
\end{gather}
\end{subequations}
of course always with $c=1$ respectively $c=1/2$ and $\tau \equiv \tau
(t) $ defined in terms of the time~$t$ by~\eqref{ztildetau}.
Note that the $N\times N$ matrix~$\tilde{A}$ is \textit{diagonal} and the $N\times N$ matrix $\tilde{B}$ is now \textit{dyadic}.
\end{proposition}

It is plain that the matrix $\tilde{U}(t) $ is \textit{%
isochronous} with period $T$ (see~(\ref{Utilde}), (\ref{ztildetau}) and~(\ref{T})),
\begin{gather}
\tilde{U} ( t\pm T ) =\tilde{U}(t),
\end{gather}
and the same property of \textit{isochrony} holds therefore for the generic
solutions of the two many-body models (\ref{NewIsoGold}), up to the
observation made above (see Remark~\ref{remark2.3}).

\section{Proofs}\label{section3}

The starting point of our treatment is the following system of two coupled
matrix ODEs satisf\/ied by the two $N\times N$ matrices $U\equiv U (t ) $ and $V\equiv V(t) $:
\begin{gather}
\dot{U}=-U^{2}+V , \qquad \dot{V}=-U V+\alpha I+\beta U+\gamma V.\label{EqUV}
\end{gather}

\begin{notation}\label{notation3.1}
The $3$ scalars $\alpha ,$ $\beta$, $\gamma $ are 3,
\textit{a~priori} arbitrary, constant parameters; $I$ is the unit $N\times N$
matrix; and we trust the rest of the notation to be self-evident (see also
Sections~\ref{section1} and~\ref{section2}).
\end{notation}

\begin{remark}\label{remark3.1}
Additional parameters could of course be introduced by
scalar shifts or rescalings of the dependent variables $U$, $V$ or of the
independent variable~$t$ (``time''). We forsake any discussion of such trivial
transformations (see Remark~\ref{remark2.2}).
\end{remark}

To solve this matrix system we introduce the $N\times N$ matrix $W\equiv
W(t) $ by setting
\begin{gather}
U(t) = [ W(t)  ] ^{-1} \dot{W} (t )  ,\qquad V(t) = [ W(t)  ] ^{-1} \ddot{W}(t).  \label{UVW}
\end{gather}
It is then easily seen that the system (\ref{EqUV}) entails that the matrix $%
W$ satisfy the following \textit{linear} \textit{third-order} matrix
ordinary dif\/ferential equation (ODE):
\begin{gather}
\dddot{W}=\alpha  W+\beta  \dot{W}+\gamma  \ddot{W};  \label{EqW}
\end{gather}
and the converse is as well true (in fact, perhaps easier to verify), namely
if $W$ satisf\/ies this \textit{linear} \textit{third-order} matrix ODE, then
the two matrices $U$ and $V$ def\/ined by~(\ref{UVW}) satisfy the system~(\ref{EqUV}).

Clearly the \textit{general} solution of this \textit{linear third-order} matrix ODE reads
\begin{gather}
W(t) =\sum_{j=1}^{3}\big[ W^{( j) }\exp (
i\omega _{j}t) \big] ,  \label{SolW}
\end{gather}%
where the $3$ \textit{constant} matrices $W^{( j) }$ are \textit{arbitrary}, $i$ is the \textit{imaginary unit} ($i^{2}=-1$; introduced here
for notational convenience), and the $3$ scalars $\omega _{j}$ are the $3$
roots of the following cubic equation in $\omega $:
\begin{subequations}
\begin{gather}
\omega ^{3}+i \gamma \omega ^{2}+\beta \omega -i\alpha =0,
\\
\omega ^{3}+i\gamma \omega ^{2}+\beta \omega -i\alpha =( \omega
-\omega _{1}) ( \omega -\omega _{2}) ( \omega -\omega
_{3}),  \label{omega123}
\end{gather}
so that
\begin{gather}
\alpha =-i \omega _{1} \omega _{2} \omega _{3} ,\qquad \beta =\omega _{1} \omega
_{2}+\omega _{2} \omega _{3}+\omega _{3} \omega _{1}, \qquad \gamma =i(
\omega _{1}+\omega _{2}+\omega _{3}) .
\end{gather}
\end{subequations}

It is then easily seen that the \textit{general} solution of the system (\ref{EqUV}) can be written as follows:
\begin{subequations}
\label{GenSolUV}
\begin{gather}
U(t) =i \big\{ I+A \exp  [ i  ( \omega _{2}-\omega
_{1} )  t ] +B \exp  [ i ( \omega _{3}-\omega _{1} )
 t ] \big\} ^{-1}  \notag \\
\hphantom{U(t) =}{}
\times \big\{ \omega _{1} I+\omega _{2} A \exp  [ i  ( \omega
_{2}-\omega _{1} )  t ] +\omega _{3} B \exp  [ i ( \omega
_{3}-\omega _{1} ) t ] \big\}  ,    \label{GenSolU}
\\
V(t) =-\big\{ I+A \exp  [ i\omega _{2}-\omega
_{1} )  t ] +B \exp  [ i ( \omega _{3}-\omega _{1} )
 t ] \big\} ^{-1}   \notag \\
\hphantom{V(t) =}{}\times
  \big\{ \omega _{1}^{2} I+\omega _{2}^{2} A \exp  [ i (
\omega _{2}-\omega _{1} )  t ] +\omega _{3}^{2} B \exp  [
i ( \omega _{3}-\omega _{1} )  t ] \big\}  ,
\label{GenSolV}
\end{gather}
\end{subequations}
where $A$ and $B$ are two, \textit{a priori} arbitrary, \textit{constant} $N\times N$ matrices. And a trivial if tedious computation shows that these
formulas provide the solution of the \textit{initial-value} problem for the
system~(\ref{EqUV}) if the two matrices $A$ and $B$ are expressed in terms
of the \textit{initial values} $U_{0}\equiv U(0)$, $V_{0}\equiv
V(0) $ as follows:
\begin{subequations}\label{SolAB}
\begin{gather}
A  = - ( \omega _{1}-\omega _{3} )  ( \omega _{2}-\omega
_{3} ) ^{-1} \big[ I-i ( \omega _{1}+\omega _{3} ) \big(
V_{0}+\omega _{3}^{2}\big) ^{-1} ( U_{0}-i \omega _{3} ) ^{-1} \big]    \notag \\
\hphantom{A=}{}\times
 \big[ I-i ( \omega _{2}+\omega _{3} ) \big(
V_{0}+\omega _{3}^{2}\big) ^{-1} ( U_{0}-i \omega _{3} ) ^{-1}\big] ^{-1},  \label{SolA}
\\
B  =  ( \omega _{1}-\omega _{2} )  ( \omega _{2}-\omega
_{3} ) ^{-1}\big[ I-i ( \omega _{1}+\omega _{2} ) \big(
V_{0}+\omega _{2}^{2}\big) ^{-1} ( U_{0}-i \omega _{2} ) ^{-1}\big]    \notag \\
\hphantom{B  =}{}
\times \big[ I-i ( \omega _{2}+\omega _{3} ) \big(
V_{0}+\omega _{2}^{2}\big) ^{-1} ( U_{0}-i \omega _{2} ) ^{-1}\big]^{-1}.  \label{SolB}
\end{gather}
\end{subequations}

To derive, from the \textit{solvable} matrix system (\ref{EqUV}), the
\textit{solvable} many-body problem reported in the preceding section, we
follow the procedure outlined in Section~\ref{section1}. This requires that we
introduce~-- in addition to the \textit{diagonal} $N\times N$ matrix $Z$
respectively the \textit{nondiagonal} $N\times N$ matrix $Y$ associated to~$U $ respectively $V$ via (\ref{UZ}) respectively~(\ref{VYR})~-- the auxiliary
$N\times N$ matrix $M\equiv M(t) $ def\/ined as follows in terms
of the \textit{diagonalizing} matrix $R(t)$, see~(\ref{UZ}):
\begin{subequations}\label{M}
\begin{gather}
M(t) = [ R(t)  ] ^{-1} \dot{R} (t )  .
\end{gather}
In the following we indicate as $\mu _{n}\equiv \mu _{n}(t) $
respectively $M_{nm}\equiv M_{nm}(t) $ the \textit{diagonal}
respectively \textit{off-diagonal} elements of this matrix:
\begin{gather}
M_{nm}=\delta _{nm} \mu _{n}+ ( 1-\delta _{nm} )  M_{nm} .
\label{Mmu}
\end{gather}
\end{subequations}

\begin{remark}\label{remark3.2}
 As implied by Remark~\ref{remark1.1}, the \textit{diagonal}
elements $\mu _{n}$ can be assigned freely, since the transformation
$R(t) \Rightarrow \tilde{R}(t) =R(t) D(t) $ with $D(t) ={\rm diag}[ d_{n}(t)] $ implies, for the \textit{diagonal} elements $\tilde{\mu}_{n}(t) $ of the matrix $\tilde{M}= [ \tilde{R} (t)] ^{-1} \overset{\cdot }{\tilde{R}}(t)$, the
expression $\tilde{\mu}_{n}(t) =\mu _{n}(t) +\dot{d}_{n}(t) /d_{n}(t) $, with a~corresponding change of
the \textit{off-diagonal} elements of the matrix $M$, $M_{nm}\Rightarrow
\tilde{M}_{nm}=\delta _{n}^{-1} M_{nm} \delta _{m}$. Note that here and
hereafter we denote as~$M_{nm}$ the \textit{off-diagonal} elements of the
matrix~$M$.
\end{remark}

It is then easily seen that the equations (\ref{UZ}) characterizing the time
evolution of $U$ and $V$ imply the following equations characterizing the
time evolution of $Z$ and $Y$:
\begin{gather}
\dot{Z}+\left[ M,Z\right] =-Z^{2}+Y,\qquad \dot{Y}+ [ M,Y]
=-ZY+\alpha I+\beta Z+\gamma Y.  \label{ZYdot}
\end{gather}

\begin{notation}\label{notation3.2}
The notation $[ A,B] $ denotes the commutator of the two matrices $A$, $B$: $[ A,B] \equiv AB-BA$.
\end{notation}

Let us now look separately at the \textit{diagonal} and \textit{off-diagonal}
parts of these two matrix equations, (\ref{ZYdot}).

The \textit{diagonal} part of the f\/irst of these two equations reads (see~(\ref{UZ}) and~(\ref{Y}))
\begin{subequations}
\begin{gather}
\dot{z}_{n}=-z_{n}^{2}+y_{n},  \label{zndot}
\end{gather}
implying%
\begin{gather}
y_{n}=\dot{z}_{n}+z_{n}^{2}.  \label{yn}
\end{gather}
\end{subequations}

Likewise, the \textit{off-diagonal} part of the f\/irst of these two equations
reads
\begin{subequations}
\begin{gather}
-( z_{n}-z_{m}) M_{nm}=Y_{nm}, \qquad n\neq m,
\end{gather}
implying
\begin{gather}
M_{nm}=-\frac{Y_{nm}}{z_{n}-z_{m}}, \qquad n\neq m.  \label{Mnm}
\end{gather}
\end{subequations}

The \textit{diagonal} part of the second of these two equations reads (see (\ref{UZ}) and (\ref{Y}))
\begin{subequations}
\begin{gather}
\dot{y}_{n}=-z_{n}y_{n}+\alpha +\beta z_{n}+\gamma y_{n}+\sum_{\ell
=1,\, \ell \neq n}^{N} ( Y_{n\ell }M_{\ell n}-M_{n\ell }Y_{\ell
n}) ,
\end{gather}
implying, via (\ref{yn}) and (\ref{Mnm}),
\begin{gather}
\dot{y}_{n}=-\dot{z}_{n} z_{n}+\gamma \dot{z}_{n}-z_{n}^{3}+\gamma
z_{n}^{2}+\beta z_{n}+\alpha +2\sum_{\ell =1,\, \ell \neq n}^{N}\left(
\frac{Y_{n\ell } Y_{\ell n}}{z_{n}-z_{\ell }}\right) .  \label{yndot}
\end{gather}
\end{subequations}

We now note that, via this equation, time-dif\/ferentiation of (\ref{zndot})
yields the following set of Newtonian-like equations of motion:
\begin{gather}
\ddot{z}_{n}=-3\dot{z}_{n}z_{n}+\gamma \dot{z}_{n}-z_{n}^{3}+\gamma
z_{n}^{2}+\beta z_{n}+\alpha +2\sum_{\ell =1,\, \ell \neq n}^{N}\left(
\frac{Y_{n\ell } Y_{\ell n}}{z_{n}-z_{\ell }}\right),  \label{Newtonlike}
\end{gather}
conf\/irming the treatment outlined in Section~\ref{section1}, see in particular (\ref{zndotdot}).

Finally we consider the \textit{off-diagonal} elements of the second of the
matrix equations~(\ref{ZYdot}). The relevant equations read, componentwise,
as follows:
\begin{subequations}
\begin{gather}
\dot{Y}_{nm}=- ( z_{n}-\gamma  )  Y_{nm}+\sum_{k=1}^{N} (Y_{nk} M_{km}-M_{nk} Y_{km} )  , \qquad n\neq m,
\end{gather}
namely, via (\ref{Y}) and (\ref{Mmu}),
\begin{gather}
\dot{Y}_{nm}=- ( z_{n}-\gamma  )  Y_{nm}- ( \mu _{n}-\mu
_{m} )  Y_{nm}+ ( y_{n}-y_{m} )  M_{nm}    \notag \\
\hphantom{\dot{Y}_{nm}=}{}
+\sum_{\ell =1,\, \ell \neq n,m}^{N} ( Y_{n\ell } M_{\ell m}-M_{n\ell
} Y_{\ell m} )  ,\qquad n\neq m.
\end{gather}
And via (\ref{yn}) and (\ref{Mnm}) (and a tiny bit of algebra) this becomes
\begin{gather}
\frac{\dot{Y}_{nm}}{Y_{nm}}=-2 z_{n}-z_{m}+\gamma -\mu _{n}+\mu _{m}-\frac{\dot{z}_{n}-\dot{z}_{m}}{z_{n}-z_{m}}   \notag \\
\hphantom{\frac{\dot{Y}_{nm}}{Y_{nm}}=}{}
+\sum_{\ell =1,\, \ell \neq n,m}^{N}\left[ \frac{Y_{n\ell } Y_{\ell m}}{Y_{nm}}\left( \frac{1}{z_{n}-z_{\ell }}+\frac{1}{z_{m}-z_{\ell }}\right) \right]
,\qquad n  \neq  m.  \label{EqYnm}
\end{gather}
\end{subequations}

The next step is to try out the \textit{ans\"{a}tze}~(\ref{Ans}), to see if
one can thereby get rid of the quantities~$Y_{nm}$.

{\sloppy We leave the (rather easy but unfortunately unproductive) task to verify
that the \textit{ansatz}~(\ref{Ans1}) does not work, i.e.\ that it does not
allow to eliminate the quantities~$Y_{nm}$ by f\/inding an assignment of the
functions $g^{(1) }(z) $ and $g^{( 2)}( z) $ which, when inserted in~(\ref{Ans1}), yield $N(N-1) $ quantities~$Y_{nm}$ that satisfy the $N( N-1) $ ODEs
(\ref{EqYnm}) (even by taking into account the possibility to assign
freely~-- see Remark~\ref{remark3.2}~-- the $N$ quantities~$\mu _{n}$).

}

We show that instead the \textit{ansatz} (\ref{Ans2}) allows the elimination
of the quantities $Y_{nm}$ and leads to the \textit{Newtonian} equations of
motion ``of goldf\/ish type''~(\ref{NewGold}). Indeed the insertion in~(\ref{EqYnm}) of (\ref{Ans2}) with
\begin{subequations}
\begin{gather}
g^{(1) }(z) =g^{(2) }(z)
=g(z)  , \qquad f^{(1) }(z) =f^{(2) }(z) =f(z),  \label{ggff}
\end{gather}
and the assignment (see Remark~\ref{remark3.2})
\begin{gather}
\mu _{n}=-\frac{z_{n}}{2}
\end{gather}
\end{subequations}
entails that the $N( N-1) $ equations (\ref{EqYnm}) can be
re-formulated as follows:
\begin{subequations}
\begin{gather}
\frac{1}{2} \left\{ \frac{\ddot{z}_{n}+\dot{z}_{n} f^{\prime } (
z_{n} ) }{\dot{z}_{n}+f(z_{n}) }+\frac{g^{\prime }(z_{n}) }{g(z_{n}) }+3z_{n}-\gamma +( (n\Rightarrow m) ) \right\} =-\frac{\dot{z}_{n}-\dot{z}_{m}}{z_{n}-z_{m}}   \notag \\
\qquad{} +\sum_{\ell =1,\, \ell \neq n,m}^{N}\left\{ g ( z_{\ell } )   [
\dot{z}_{\ell }+f ( z_{\ell } )  ]  \left( \frac{1}{z_{n}-z_{\ell }}+\frac{1}{z_{m}-z_{\ell }}\right) \right\} ,\qquad  n\neq m.
\end{gather}

\begin{notation}\label{notation3.3}
Here and below primes indicate dif\/ferentiations with
respect to the argument of the functions they are appended to; and the
convenient shorthand notation $+( ( n\Rightarrow m) ) $
denotes addition of whatever comes before it, with the index~$n$ replaced by
the index~$m$.
\end{notation}

It is easily seen that these equations can be re-written as follows:%
\begin{gather}
\frac{\ddot{z}_{n}+\dot{z}_{n} f^{\prime }(z_{n}) }{\dot{z}_{n}+f(z_{n}) }+\frac{g^{\prime }(z_{n}) }{g(
z_{n}) }+3z_{n}-\gamma -\sum_{\ell =1,\, \ell \neq n}^{N}\left\{ \frac{g( z_{\ell }) [ \dot{z}_{\ell }+f( z_{\ell })
] }{z_{n}-z_{\ell }}\right\} +( ( n\Rightarrow m))  \label{zdotdotnm} \\
\qquad{} =2\left\{ \frac{[ g(z_{n}) -1] \dot{z}_{n}-[
g( z_{m}) -1] \dot{z}_{m}+g(z_{n}) f(z_{n}) -g( z_{m}) f( z_{m}) }{z_{n}-z_{m}}\right\} ,\qquad n\neq m.\notag
\end{gather}
\end{subequations}

This suggests the assignments
\begin{gather}
g(z) =1,\qquad f(z) =a+bz+cz^{2},  \label{gf}
\end{gather}
with the $3$ parameters $a$, $b$, $c$ \textit{a priori arbitrary}, since thereby
the $N ( N-1 ) $ equations~(\ref{zdotdotnm}) get reduced to the~$N$
equations
\begin{subequations}
\begin{gather}
\frac{\ddot{z}_{n}+\dot{z}_{n} ( b+2cz_{n} ) }{\dot{z}_{n}+a+bz_{n}+cz_{n}^{2}}=( 2c-3 )  z_{n}+\gamma
+b+2 \sum_{\ell =1,\, \ell \neq n}^{N}\left( \frac{\dot{z}_{\ell }+a+bz_{\ell
}+cz_{\ell }^{2}}{z_{n}-z_{\ell }}\right),
\end{gather}
or equivalently (in Newtonian form)
\begin{gather}
\ddot{z}_{n}=-3\dot{z}_{n}z_{n}+\gamma \dot{z}_{n}+ ( 2c-3 )
cz_{n}^{3}+[ \gamma c+3b( c-1) ] z_{n}^{2}
\notag \\
\hphantom{\ddot{z}_{n}=}{}
+ [ a( 2c-3) +b ( \gamma +b )  ]
 z_{n}+a ( \gamma +b )    \notag \\
\hphantom{\ddot{z}_{n}=}{}
+\sum_{\ell =1,\, \ell \neq n}^{N}\left[ \frac{2\big( \dot{z}
_{n}+a+bz_{n}+cz_{n}^{2}\big)  \big( \dot{z}_{\ell }+a+bz_{\ell
}+cz_{\ell }^{2}\big) }{z_{n}-z_{\ell }}\right] .
\end{gather}
\end{subequations}

Consistency requires now that this set of $N$ Newtonian equations of motions
coincide with the $N$ analogous equations (\ref{Newtonlike}), which, via (\ref{Ans2}) with (\ref{ggff}) and (\ref{gf}), now read
\begin{gather}
\ddot{z}_{n}=-3 \dot{z}_{n} z_{n}+\gamma \dot{z}_{n}-z_{n}^{3}+\gamma
z_{n}^{2}+\beta z_{n}+\alpha   \notag \\
\hphantom{\ddot{z}_{n}=}{}
+\sum_{\ell =1,\, \ell \neq n}^{N}\left[ \frac{2  \big( \dot{z}_{n}+a+bz_{n}+cz_{n}^{2}\big) \big( \dot{z}_{\ell }+a+bz_{\ell
}+cz_{\ell }^{2}\big) }{z_{n}-z_{\ell }}\right].
\end{gather}
This clearly requires that the following $4$ constraints on the $6$
parameters $\alpha$, $\beta$, $\gamma$, $a$, $b$, $c$ be~satisf\/ied (note
that, somewhat miraculously, the two velocity-dependent one-body terms in
the right-hand sides of the last two equations match automatically, as well
as the two-body terms):
\begin{subequations}\label{Constraints}
\begin{gather}
( 2c-3) c=-1,  \label{Eqc}
\\
\gamma c+3b( c-1) =\gamma ,  \label{Eqcgamma}
\\
a( 2c-3) +b( \gamma +b) =\beta ,
\\
a( \gamma +b) =\alpha .
\end{gather}
\end{subequations}
And it is easily seen that this entails two alternative possibilities:

\textit{model $(i)$}: $a$, $b$, $\gamma $ arbitrary and
\begin{subequations}
\begin{gather}
c=1, \qquad \beta =-a+b ( \gamma +b ) ,\qquad \alpha =a ( \gamma
+b ) ;  \label{alphabeta1}
\end{gather}

\textit{model $(ii)$}: $a$, $b$ arbitrary and
\begin{gather}
c=\frac{1}{2}, \qquad \gamma =-3b,\qquad \beta =-2 \big( a+b^{2}\big)
 ,\qquad \alpha =-2ab.  \label{alphabetagamma2}
\end{gather}
\end{subequations}
(Note that, in \textit{case (i)}, another miracle occurred: the solution $c=1 $ of the f\/irst,~(\ref{Eqc}), of the $4$ constraints (\ref{Constraints})
entailed that the second,~(\ref{Eqcgamma}), of these $4$ equations hold
identically).

Clearly these two possibilities correspond to the two \textit{solvable}
many-body models ``of goldf\/ish type'' (\ref{NewGold}).

Next, we must justify the assertions made in the preceding section (see
Proposition~\ref{proposition2.1}) concerning the solution of the initial-value
problems for the many-body models characterized by the Newtonian equations
of motion~(\ref{NewGold}). The treatment given above (in this section)
entails that these solutions are provided by the eigenvalues of the $N\times
N$ matrix $U(t) $ evolving according to the explicit formula~(\ref{GenSolU}) with~(\ref{SolAB}); the missing detail is to express the two
initial $N\times N$ matrices $U_{0}\equiv U(0) $ and $V_{0}\equiv V(0) $ appearing in the right-hand side of~(\ref{SolAB}) in terms of the $2N$ initial data, $z_{n}(0) $ and $\dot{z}_{n}(0) $, of the many-body problems~(\ref{NewGold}).

To simplify the derivation of these formulas it is convenient to make the
assumption (allowed by the treatment given above) that the matrix $U(t) $ be \textit{initially diagonal}, namely (see~(\ref{UZ})) that
\begin{gather}
R(0) =I.  \label{Rzero}
\end{gather}
This entails (see (\ref{UZ}) and (\ref{VYR})) that
\begin{gather}
U(0) =\text{diag} [ z_{n}(0)  ] ,\qquad V(0) =Y(0).  \label{UVzero}
\end{gather}

The formula (\ref{Uzero}) for $U_{0}$ is thereby immediately implied.

The formula (\ref{Vzero}) for $V_{0}\equiv V(0) =Y(0) $ is also easily obtained, since the expression~(\ref{yn}) of the
\textit{diagonal} elements of the matrix $Y$ and the \textit{ansatz}~(\ref{Ans2}) with~(\ref{ggff}) and~(\ref{gf}) for the \textit{off-diagonal}
elements of $Y$ entail that, componentwise,
\begin{gather}
Y_{nm}(0) =\delta _{nm}  [ \dot{z}_{n}(0)
+z_{n}^{2}(0) ] +( 1-\delta _{nm}) \big[ \dot{z}_{n}(0)
+a+bz_{n}(0) +cz_{n}^{2}(0) \big] ^{1/2}
\notag \\
\hphantom{Y_{nm}(0) =}{}
\times \big[ \dot{z}_{m}(0) +a+bz_{m}(0)
+cz_{m}^{2}(0) \big] ^{1/2},
\end{gather}
and this clearly yields (\ref{Vzero}).

This completes the proof of the results of the preceding Section~\ref{section2}. As for
the f\/indings reported in Section~\ref{section2.1}, we consider their derivation
suf\/f\/iciently obvious~-- for instance via the treatment detailed in Section~2.1 of~\cite{C2008}; and by repeating the relevant treatment as given above,
especially in the last part of this section~-- to justify us to dispense here
from any further elaboration.

\section{Additional f\/indings}\label{section4}

In this section we introduce the time-dependent (monic) polynomial $\psi( z,t) $ whose zeros are the $N$ eigenvalues $z_{n}(t) $ of the $N\times N$ matrix $U(t)$:
\begin{subequations}
\label{psi}
\begin{gather}
\psi  ( z,t ) =\det  [ zI-U(t)  ] ,
\\
\psi  ( z,t ) =\prod\limits_{n=1}^{N} [ z-z_{n}(t) ] =z^{N}+\sum_{m=1}^{N}\big[ c_{m}(t)  z^{N-m}\big].
\end{gather}
The last of these formulas introduces the $N$ coef\/f\/icients $c_{m}\equiv
c_{m}(t) $ of the monic polyno\-mial~$\psi ( z,t )$; of
course it implies that these coef\/f\/icients are related to the zeros $z_{n}(t) $ as follows:
\begin{gather}
c_{1}=-\sum_{n=1}^{N}z_{n}, \qquad c_{2}=\sum_{n,m=1,\, n>m}^{N}z_{n}z_{m},
\label{c12}
\end{gather}
\end{subequations}
and so on.

The fact that the initial-value problem associated with the time evolution (\ref{NewGold}) of the~$N$ coordinates~$z_{n}$ can be \textit{solved} by
\textit{algebraic} operations implies that the same \textit{solvable}
character holds for the time evolution of the monic polynomial $\psi (z,t) $ and of the $N$ coef\/f\/icients~$c_{m}(t) $. In this
section we display explicitly the equations that characterize these time
evolutions. The procedure to obtain these equations from the equations of
motion~(\ref{NewGold}) is quite tedious but standard; a key role in this
development are the identities reported, for instance, in Appendix~A of \cite{C2008} (but there are 2 misprints in these formulas: in equation~(A.8k) the term
$( N+1) $ inside the square bracket should instead read $(N-3) $; in equation~(A.8l) the term $N^{2}$ inside the square bracket
should instead read $N( N-2)$). Here we limit our presentation
to reporting the f\/inal result.

The equation characterizing the time evolution of the monic polynomial $\psi
( z,t) $ implied by the Newtonian equations of motion (\ref{NewGold}) reads as follows:
\begin{subequations}
\begin{gather}
 \psi _{tt}+ \big( \eta _{0}+\eta _{1} z+\eta _{2} z^{2} \big)  \psi
_{zt}+ ( \theta _{0}+\theta _{1} z )  \psi _{t}   \notag \\
\qquad{} +\left( \alpha _{0}+\alpha _{1} z+\alpha _{2} z^{2}+\alpha
_{3} z^{3}+\alpha _{4}z^{4}\right) \psi _{zz}+\big( \beta _{0}+\beta
_{1}z+\beta _{2}z^{2}+\beta _{3}z^{3}\big)  \psi _{z}   \notag \\
\qquad{} +\big( \gamma _{0}+\gamma _{1}z+\gamma _{2}z^{2}\big) \psi =0.
\label{Eqpsi}
\end{gather}
Here the subscripted variables denote partial dif\/ferentiations, and the $17$
coef\/f\/icients appearing in this equation are def\/ined in terms of the
quantities $c_{1}\equiv c_{1}(t)$, $\dot{c}_{1}\equiv \dot{c}_{1}(t) $ and $c_{2}\equiv c_{2}(t) $, see~(\ref{c12}), of the \textit{arbitrary positive integer} $N$ and of the $3$ free
parameters~$a$,~$b$ and $\gamma $ characterizing \textit{model $(i)$},
respectively of the $2$ free parameters $a$ and $b$ characterizing \textit{model $(ii)$}, as follows:

\textit{model $(i)$}:
\begin{gather}
\eta _{0} =-2a,\qquad \eta _{1}=-2b, \qquad \eta _{2}=-2;  \notag \\
\theta _{0} =-2c_{1}+2\left( N-1\right) b-\gamma,\qquad \theta
_{1}=2 N-1;  \notag \\
\alpha _{0} = a^{2} ,\qquad \alpha _{1}=2ab, \qquad \alpha
_{2}=2a+b^{2},\qquad \alpha _{3}=2b,\qquad \alpha _{4}=1;  \notag \\
\beta _{0} =[ 2c_{1}-( 2 N-3) b+\gamma ]
a,\qquad \beta _{1}=2bc_{1}- ( 2 N-3 ) \big( a+b^{2}\big)
+b\gamma ,  \notag \\
\beta _{2} =2c_{1}-2( 2N-3) b+\gamma, \qquad \beta
_{3}=- ( 2N-3);  \notag \\
\gamma _{0} =-\dot{c}_{1}+c_{1}^{2}-2( N-1) bc_{1}+\gamma
c_{1}-Na+N( N-2) b^{2}-Nb\gamma ,  \notag \\
\gamma _{1} =-( 2N-1) c_{1}+2N( N-2) b-N\gamma
,\qquad \gamma _{2}=N ( N-2 ) ;
\end{gather}

\textit{model $(ii)$}:
\begin{gather}
\eta _{0}  = -2a,\qquad \eta _{1}=-2b,\qquad \eta _{2}=-1;  \notag \\
\theta _{0} =-c_{1}+2( N+1) b,\qquad \theta _{1}=N+1;  \notag \\
\alpha _{0} =a^{2},\qquad \alpha _{1}=2ab,\qquad \alpha
_{2}=a+b^{2}, \qquad \alpha _{3}=b, \qquad \alpha _{4}=\frac{1}{4};  \notag \\
\beta _{0} =( c_{1}-2Nb) a, \qquad \beta
_{1}=bc_{1}-Na-2Nb^{2},  \qquad
\beta _{2} =\frac{c_{1}}{2}-2Nb, \qquad \beta _{3}=-\frac{N}{2};  \notag \\
\gamma _{0} =-2\dot{c}_{1}+c_{1}^{2}- ( N+2 ) bc_{1}-\frac{3c_{2}}{2}+Na+N( N+1) b^{2},  \notag \\
\gamma _{1} =-\frac{( N+1) c_{1}}{2}+N( N+1)
b, \qquad \gamma _{2}=\frac{N ( N+1) }{4}.
\end{gather}
\end{subequations}

\begin{remark}\label{remark4.1}
The equation~(\ref{Eqpsi}) characterizing the time
evolution of the monic polynomial~$\psi $ looks like a \textit{linear
partial differential equation}, but it is in fact a \textit{nonlinear
functional equation}, because some of its coef\/f\/icients depend on the
quantities $c_{1}$ and $c_{2}$ which themselves depend on~$\psi $, indeed
clearly (see~(\ref{psi}))
\begin{subequations}
\begin{gather}
c_{1}\equiv c_{1}(t) =\frac{\psi ^{( N-1) }(
0,t) }{( N-1) !},\qquad c_{2}\equiv c_{2}(t) =
\frac{\psi ^{( N-2) }( 0,t) }{( N-2) !},
\end{gather}
where we used the shorthand notation $\psi ^{( j) }(z,t) $ to indicate the $j$-th partial derivative with respect to the
variable $z$ of $\psi ( z,t) $,
\begin{gather}
\psi ^{( j) }( z,t) \equiv \frac{\partial ^{j}\psi
( z,t) }{\partial z^{j}},\qquad j=1,2,\dots.
\end{gather}
\end{subequations}
\end{remark}

As for the system of $N$ nonlinear autonomous second-order ODEs of \textit{Newtonian} type cha\-racterizing the time evolution of the $N$ coef\/f\/icients~$c_{m}$, for \textit{model $(i)$} they read as follows:
\begin{subequations}
\label{Eqcm1}
\begin{gather}
 \ddot{c}_{m}+p_{m}^{(-1) } \dot{c}_{m-1}+p_{m}^{ (0) }\dot{c}_{m}+p_{m}^{(1) }\dot{c}_{m+1}  \notag \\
\qquad \quad{} +q_{m}^{(-2) }c_{m-2}+q_{m}^{(-1)
}c_{m-1}+q_{m}^{(0) }c_{m}+q_{m}^{(1)
}c_{m+1}+q_{m}^{(2) }c_{m+2}  \notag \\
\qquad{} =2c_{1}\dot{c}_{m}-2( N-m+1) ac_{1}c_{m-1}  \notag \\
\qquad\quad{} +\big[ \dot{c}_{1}-c_{1}^{2}+2 ( m-1 ) bc_{1}-\gamma c_{1}\big] c_{m}+ ( 2m+1 ) c_{1}c_{m+1},  \label{Eqc1}
\end{gather}
with the $3N$ (time-independent) coef\/f\/icients $p_{m}^{( j) }$ and
the $5N$ (time-independent) coef\/f\/i\-cients~$q_{m}^{( j) }$ def\/ined
here in terms of the $3$ \textit{free} parameters $a$, $b$ and $\gamma $ (and
of the numbers~$m$ and~$N$) as follows:
\begin{gather}
p_{m}^{(-1) }  = -2 ( N-m+1 ) a, \qquad p_{m}^{(
0) }=2( m-1) b-\gamma ,\qquad p_{m}^{(1)
}=2m+1;  \notag \\
q_{m}^{(-2) } =( N-m+2) ( N-m+1)a^{2},  \notag \\
q_{m}^{(-1) } =( N-m+1) [ -(2m-3) b+\gamma ] a,  \notag \\
q_{m}^{(0) } =-m( 2N-2m+1) a+m(
m-2) b^{2}-mb\gamma ,  \notag \\
q_{m}^{(1) } =2\big( m^{2}-1\big) b-( m+1)
\gamma,\qquad q_{m}^{(2) }=m ( m+2 ) .
\end{gather}
\end{subequations}

The analogous equations for \textit{model $(ii)$} read as follows:
\begin{subequations}\label{Eqcm2}
\begin{gather}
 \ddot{c}_{m}+p_{m}^{(-1) }\dot{c}_{m-1}+p_{m}^{(0) }\dot{c}_{m}+p_{m}^{(1) }\dot{c}_{m+1} \notag\\
 \qquad\quad{}
 +q_{m}^{(-2) }c_{m-2}+q_{m}^{(-1)
}c_{m-1}+q_{m}^{(0) }c_{m}+q_{m}^{(1)
}c_{m+1}+q_{m}^{(2) }c_{m+2}  \notag \\
\qquad{}=c_{1}\dot{c}_{m}- ( N-m+1 ) ac_{1}c_{m-1}  \notag \\
\qquad\quad{}+\left[ 2\dot{c}_{1}+\frac{3c_{2}}{2}-c_{1}^{2}+ ( m+2 )
bc_{1}\right] c_{m}
+\left( \frac{m}{2}+1\right) c_{1}c_{m+1},  \label{Eqc2}
\end{gather}
with the $3N$ (time-independent) coef\/f\/icients $p_{m}^{(j) }$ and
the $5N$ (time-independent) coef\/f\/i\-cients~$q_{m}^{( j) }$ def\/ined
here in terms of the $2$ \textit{free} parameters $a$ and $b$ (and of the
numbers~$m$ and~$N$) as follows:
\begin{gather}
p_{m}^{(-1) }  = -2 ( N-m+1 ) a, \qquad p_{m}^{(0) }=2 ( m+1 ) b, \qquad p_{m}^{(1) }=m+2;  \notag
\\
q_{m}^{(-2) } =( N-m+2) ( N-m+1)
a^{2}, \qquad q_{m}^{(-1) }=-2m( N-m+1) ab,  \notag
\\
q_{m}^{(0) } =-m( N-m-1) a+m( m+1)
b^{2},  \notag \\
q_{m}^{(1) } =( m+1) ( m+2)
b, \qquad q_{m}^{(2) }=\frac{( m+2) ( m+3) }{4}.
\end{gather}
\end{subequations}

Of course in these equations of motion, (\ref{Eqc1}) and (\ref{Eqc2}), it is
understood that, for $n<0$ and for $n>N$, the coef\/f\/icients~$c_{n}$ vanish
identically, $c_{n}=0$, while $c_{0}=1$ (see~(\ref{psi})).

It is plain that these equations of motion,~(\ref{Eqcm1}) and~(\ref{Eqcm2}),
inherit the properties of the original many-body models: they clearly are as
well \textit{solvable} by algebraic operations, and of course if the
original many-body model is \textit{isochronous} the corresponding model for
the coef\/f\/icients~$c_{m}$ is as well \textit{isochronous}, namely
\begin{subequations}
\begin{gather}
c_{m}( t\pm T) =c_{m}(t),
\end{gather}
and likewise if the original many-body model is \textit{asymptotically
isochronous}, the corresponding model for the coef\/f\/icients~$c_{m}$ is as
well \textit{asymptotically isochronous}, namely
\begin{gather}
\lim_{t\rightarrow \infty}[ c_{m}( t\pm T)-c_{m}(t) ] =0.
\end{gather}
This of course entails that the conditions on the $3$ \textit{a priori free}
parameters~$a$,~$b$ and $\gamma $ of the version (\ref{Eqcm1}) of \textit{model~$(i)$}, or on the $2$ \textit{a priori free} parameters~$a$ and~$b$ of
the version~(\ref{Eqcm2}) of \textit{model~$(ii)$}, which are necessary and
suf\/f\/icient to imply that these models be \textit{isochronous} respectively
\textit{asymptotically isochronous}, are those indicated in Section~\ref{section1},
namely are those implied by~(\ref{cc}) with~(\ref{alphabetagammaomega}) and~(\ref{omegaintegers}) respectively (\ref{Reomegaintegers}); while in the
\textit{generic} case, see~(\ref{Imomegajdiff}), the coef\/f\/icients $c_{m}(t) $ tend asymptotically to time-independent values,
\begin{gather}
\lim_{t\rightarrow \pm \infty}[ c_{m}(t) ]
=c_{m}( \pm \infty ).
\end{gather}
\end{subequations}

\subsection{Two additional \textit{isochronous} models}\label{section4.1}

Here we report formulas analogous to those reported above, but related to
the many-body models discussed in Section~\ref{section2.1} rather than to those reported
in Section 2. We consider the derivation of these results suf\/f\/iciently
straightforward not to require any detailed elaboration here beyond the
terse hints provided below.

The starting point are the two special cases of the two systems~(\ref{Eqc1})
and~(\ref{Eqc2}) which obtain by setting all the free parameters to vanish.
We write them in compact form as follows:
\begin{gather}
\ddot{c}_{m}  = 2 c c_{1} \dot{c}_{m}- [ 2c( m-1) +3]
\dot{c}_{m+1}  +\big[  ( 3-2c ) \dot{c}_{1}-c_{1}^{2}+2\big( 1-c^{2}\big)
 c_{2}\big]  c_{m}  \notag \\
\hphantom{\ddot{c}_{m}  =}{}
+\big( 2c^{2}m+1\big) c_{1}c_{m+1}-( m+2) \big[
c^{2} ( m-1 ) +1\big] c_{m+2}
\end{gather}
with $c=1$ respectively $c=1/2$. It is remarkable that the number $N$ does
not appear in these equations; although we always assume that these
equations hold for $m=1,2,\dots,N$ and that the dependent variables $c_{n}$
vanish identically for $n>N$, with $N$ an \textit{arbitrary positive integer}. The diligent reader may also check that this equation also holds
identically for~$m=0$ with~$c_{0}=1$ (see~(\ref{psi})).

We then make the following change of dependent and independent variables:
\begin{gather}
\tilde{c}_{m}(t) =\exp ( i m\omega t)  c_{m}(
\tau ) ,\qquad \tau =\frac{\exp  ( i\omega t) -1}{i\omega },
\label{cmtilde}
\end{gather}
with $\omega $ again a \textit{real} arbitrary constant to which we
associate the period $T$ (see~(\ref{T})). One thereby easily sees that the
new dependent variables $\tilde{c}_{m}(t) $ satisfy the
following system of \textit{autonomous} Newtonian equations of motion:
\begin{gather}
\overset{\cdot \cdot }{\tilde{c}}_{m}  =  \big[  ( 2m+1 )
 i \omega +2c\tilde{c}_{1}\big]  \overset{\cdot }{\tilde{c}}_{m}- [
2c( m-1) +3] \overset{\cdot }{\tilde{c}}_{m+1}  \notag \\
\hphantom{\overset{\cdot \cdot }{\tilde{c}}_{m}  =}{}
+\big\{ m ( m+1 )  \omega ^{2}- [ 2 c ( m-1 ) +3] i\omega \tilde{c}_{1}
 + ( 3-2 c )  \overset{\cdot }{\tilde{c}}_{1}-\tilde{c}
_{1}^{2}+2\big( 1-c^{2}\big)  \tilde{c}_{2}\big\} \tilde{c}_{m}
\notag \\
\hphantom{\overset{\cdot \cdot }{\tilde{c}}_{m}  =}{}
+\big\{  ( m+1 )  [ 2c( m-1) +3]
i\omega +\big( 2c^{2} m+1\big) \tilde{c}_{1}\big\} \tilde{c}_{m+1}
\notag \\
\hphantom{\overset{\cdot \cdot }{\tilde{c}}_{m}  =}{}
- ( m+2 ) \big[ c^{2} ( m-1 ) +1\big] \tilde{c}_{m+2}.  \label{Eqcmtilde}
\end{gather}

It is plain from the way these two models (with $c=1$ or $c=1/2)$ have been
derived that they are \textit{isochronous}, namely the generic solutions of
these nonlinear Newtonian equations of motion satisfy the condition%
\begin{gather}
\tilde{c}_{m} ( t+T ) =\tilde{c}_{m}(t).
\end{gather}

\section{Outlook}\label{section5}

It is clearly far from trivial that the Newtonian many-body models
introduced in this paper~-- see~(\ref{NewGold}),~(\ref{NewIsoGold}),~(\ref{Eqcm1}),~(\ref{Eqcm2}), and~(\ref{Eqcmtilde})~-- can be \textit{solved}, for
arbitrary initial data, by \textit{algebraic} operations. Also remarkable is
that, for special assignments of their parameters, the systems of \textit{autonomous} \textit{nonlinear} ODEs~(\ref{NewGold}),~(\ref{Eqcm1}) and~(\ref{Eqcm2}) are \textit{isochronous} or \textit{asymptotically isochronous},
and the systems of \textit{autonomous} \textit{nonlinear} ODEs~(\ref{NewIsoGold}) and~(\ref{Eqcmtilde}) are \textit{isochronous}.

Let us end this paper by pointing out that \textit{Diophantine} f\/indings can
be obtained from a~\textit{nonlinear} \textit{autonomous} \textit{isochronous} dynamical system by investigating its behavior in the \textit{infinitesimal vicinity} of its equilibria. The relevant equations of motion
become then generally \textit{linear}, but they of course retain the
properties to be \textit{autonomous} and \textit{isochronous}. For a system
of \textit{linear} \textit{autonomous} ODEs, the property of \textit{isochrony} implies that \textit{all} the eigenvalues of the matrix of its
coef\/f\/icients are \textit{integer numbers} (up to a common rescaling factor).
When the \textit{linear} system describes the behavior of a \textit{nonlinear autonomous} system in the \textit{infinitesimal vicinity} of its
equilibria, these matrices can generally be \textit{explicitly} computed in
terms of the values at equilibrium of the dependent variables of the
original, \textit{nonlinear} model. In this manner nontrivial \textit{Diophantine} f\/indings and conjectures have been discovered and proposed: see
for instance the review of such developments in Appendix~C (entitled
``Diophantine f\/indings and conjectures'') of~\cite{C2008}. Analogous results
obtained by applying this approach to the \textit{isochronous} systems of
\textit{autonomous nonlinear} ODEs introduced above will be reported in a
separate paper if they turn out to be novel and interesting.

\pdfbookmark[1]{References}{ref}
\LastPageEnding

\end{document}